# Multi-temperature atomic ensemble: nonequilibrium evolution after ultrafast electronic excitation

Nikita Medvedev[1,2,*], Alexander E. Volkov[3]

1. Institute of Physics of the Czech Academy of Sciences, Na Slovance 1999/2, 182 21 Prague 8, Czech Republic
2. Institute of Plasma Physics of the Czech Academy of Sciences, Za Slovankou 1782/3, 182 00 Prague 8, Czech Republic
3. P.N. Lebedev Physical Institute of the Russian Academy of Sciences, Leninskij pr., 53,119991 Moscow, Russia

## Abstract

Ultrafast laser radiation or beams of fast charged particles primarily excite the electronic system of a solid driving the target transiently out of thermal equilibrium. Apart from the nonequilibrium between the electrons and atoms, each subsystem may be far from equilibrium. From the first principles, we derive the definition of various atomic temperatures applicable to electronically excited ensembles. It is shown that the definition of the kinetic temperature of atoms in the momentum subspace is unaffected by the excitation of the electronic system. When the electronic temperature differs from the atomic one, an expression for the configurational atomic temperature is proposed, applicable to the electronic-temperature-dependent interatomic potentials (such as *ab-initio* molecular dynamics simulations). We study how the configurational temperature behaves during nonthermal phase transition, triggered by the evolution of the interatomic potential due to the electronic excitation. It is revealed that upon the ultrafast irradiation, the atomic system of a solid exists temporarily in a multi-temperature state: separate equilibria in the momentum and configurational subspaces. Complete equilibration between the various atomic temperatures takes place at longer timescales, forming the energy equipartition. Based on these results, we propose a formulation of multi-temperature heat transport equations.

[*] Email: nikita.medvedev@fzu.cz

# I. Introduction

Systems under intensive energy flows often exhibit highly nonequilibrium and unusual processes. In particular, they occur during irradiation of matter with ultrafast laser pulses or beams of swift charged particles. In such scenarios, the energy is absorbed by target electrons, leading to the transiently strongly out-of-equilibrium material state, in which the electronic system is highly excited whereas the atoms initially are still in the cold state [1–4].

After the energy deposition, electrons redistribute their energy, thermalizing to an equilibrium distribution with a temperature high above the atomic one [1,3,5]. The atoms then respond to both: a change in the interatomic potential due to the electronic excitation (because electrons participate in the formation of the interatomic potential in matter), and the energy transfer due to scattering of the hot electrons [6]. The atomic ensemble, responding to the effects caused by the electronic excitation, may also depart from its equilibrium state. E.g. various collective atomic modes (phonon oscillations) may not be in equilibrium among themselves during and after irradiation [7,8]. Subsequent co-evolution of the highly nonequilibrium atomic and electronic ensembles leads to material equilibration at longer timescales.

The standard simulation methods of the irradiation problem include molecular dynamics (MD), often combined with the two-temperature model to account for the effect of the electronic excitation followed by energy transport and exchange between electrons and atoms [1,5]. Interatomic potentials for such simulations include pair-potentials, many-body potentials, or *ab-initio*-based simulations such as tight-binding or density-functional-theory MD [9–12]. To model ultrafast laser or ion beam irradiation problems, interatomic potentials that depend on the electronic temperature have been actively developed in the past decade [13–17].

In MD simulations, whether with classical or *ab-initio*-based potentials, it is still an open problem – how to assess the degree of disequilibrium of an atomic ensemble [18]. Generally speaking, the equilibrium state assumes the maximization of the coarse-grained entropy in the system of simulated ensembles of particles [19,20]. However, in practice, it is hard to access such a quantity in MD simulations taking into account the contribution of the distribution of atoms by the potential energy (although approximate methods exist [21,22]).

To evaluate the equilibration in atomistic simulations, several methods were developed relying on the concept of nonequilibrium temperatures, such as Rugh's temperature or configurational temperature [18,23–26]. The approach introduces a few partial temperatures related to different dynamical and configuration modes of the ensemble being investigated and compares their values during the ensemble evolution. It is assumed that the equilibrium is achieved when these temperatures start to coincide forming the unique thermodynamic temperature.

So far, this method has not been extensively developed for practical use in simulations and has not found its deserved widespread application. A part of the reason for this is that it is difficult in practice to calculate the configurational or Rugh's temperatures for realistic MD potentials. Only pair-potentials have been tested thus far [18]. Another reason is that there was no systematic study and demonstration of the importance of various temperature definitions in physically relevant nonequilibrium scenarios, such as the modeling of ultrafast laser irradiation of materials. The influence of electrons on the equilibration in the atomic ensemble has also not been studied before.

We demonstrate that the standard definitions of the configurational temperature are only applicable in static (independent of the electronic state) interatomic potentials used in standard MD simulations. The presented generalization of the definition of the various atomic configurational temperatures includes the dependence on the electronic temperature. Applying this definition, we simulate the ultrafast laser irradiation of materials, demonstrating different equilibration processes taking place at different timescales. We discuss the transient multi-temperature state forming in the nonequilibrium electronic-atomic system, which paves the way to a plausible formulation of nonequilibrium thermodynamics of the relaxation process.

## II. Theory

### 1. Preliminary considerations of the generalized temperature

Following a few pioneering works, it was realized that a temperature may be defined in a very general way *via* an arbitrary vector field in the phase space $\{\vec{R}, \vec{P}\}$ [18,24,26]. Illustrating the idea, let us consider a single-particle phase space and Gibbs distribution. From the definition of an ensemble average, the following holds for an arbitrary vector field $\boldsymbol{B}(\vec{R}, \vec{P})$ with very general constraints ($|\langle \boldsymbol{B}(\vec{R}, \vec{P})\rangle| < \infty$ and $0 < |\langle \boldsymbol{\nabla} \boldsymbol{B}(\vec{R}, \vec{P})\rangle| < \infty$; for a detailed and rigorous derivation, see Ref. [26]):

$$\begin{aligned}
\int_\Omega \boldsymbol{\nabla} \left[\boldsymbol{B}(\vec{R}, \vec{P}) \exp\left(-\frac{H}{T}\right)\right] d\vec{R} d\vec{P} &= 0 \\
&= \int_\Omega \left(\boldsymbol{\nabla} \cdot \boldsymbol{B}(\vec{R}, \vec{P})\right) \exp\left(-\frac{H}{T}\right) d\vec{R} d\vec{P} \\
&+ \int_\Omega \boldsymbol{B}(\vec{R}, \vec{P}) \cdot \left(\boldsymbol{\nabla} \exp\left(-\frac{H}{T}\right)\right) d\vec{R} d\vec{P} \\
&= \langle \boldsymbol{\nabla} \cdot \boldsymbol{B}(\vec{R}, \vec{P})\rangle - \frac{1}{T}\langle \boldsymbol{B}(\vec{R}, \vec{P}) \cdot \boldsymbol{\nabla} H\rangle = 0,
\end{aligned} \quad (1)$$

where *H* is the Hamiltonian of the ensemble of particles, *T* is the temperature – the parameter of the corresponding Gibbs distribution (here and further, the Boltzmann constant is set $k_B = 1$, *i.e.* all the temperatures are in energy units), the integration is carried over the entire phase space *Ω*, the nabla operator is for differentiation over phase variables, and the angle brackets denote an ensemble average.

Eq.(1) can be used to define the generalized temperature in a one-component system corresponding to the field $\boldsymbol{B}(\vec{R}, \vec{P})$ and an arbitrary (nonequilibrium) distribution as follows:

$$T = \frac{\langle \boldsymbol{B}(\vec{R}, \vec{P}) \cdot \boldsymbol{\nabla} H\rangle}{\langle \boldsymbol{\nabla} \cdot \boldsymbol{B}(\vec{R}, \vec{P})\rangle}. \quad (2)$$

Applying different $\boldsymbol{B}(\vec{R},\vec{P})$ in the definition (2), one finds various generalized temperatures in the arbitrary, nonequilibrium ensemble [18]. In the thermodynamic equilibrium (Gibbs distribution), all the temperatures defined for any field $\boldsymbol{B}(\vec{R},\vec{P})$ reduce to the unique thermodynamic temperature.

Generally speaking, there are three classes of single-particle vector fields $\boldsymbol{B}(\vec{R},\vec{P})$: those over only the momentum space $\boldsymbol{B}(\vec{P})$, those over only the configurational (coordinate) space $\boldsymbol{B}(\vec{R})$, and those depending on both phase variables (such as Rugh's temperature, for instance [18,23]). The real-space and the momentum-space fields are very convenient for simulations and applications, because they allow for a straightforward interpretation as particle distributions in the momentum and configurational subspaces, correspondingly. Below, we will consider these classes in more detail.

However, the definition in Eq.(2) is only applicable to a one-component system, such as an atomic ensemble interacting through a classical potential dependent only on the parameters of the atoms in the ensemble. When electrons are considered, the definition of the generalized temperatures must be modified.

## 2. Generalized temperature in a two-temperature state

If we consider an ultrafast irradiation scenario, the irradiated matter may transiently be in a two-temperature state: the electronic temperature is different from the atomic one [1,2]. In this case, the definition of the generalized atomic temperature may be derived following the same recipe as in Eq.(1), but taking into account that the distribution function of the entire system (ensemble of strongly-interacting ions and electrons) is transiently factorized into atomic and electronic equilibrium terms forming two-temperature state:

$$f = A f_i^{eq} f_e^{eq} = A\, exp\left(-\frac{K_i + U_i + \frac{1}{2}V_{ei}}{T_i}\right) exp\left(-\frac{K_e + U_e + \frac{1}{2}V_{ei}}{T_e}\right). \qquad (3)$$

Here, the total energy (Hamiltonian) of the system is decomposed into the ionic and electronic contributions:

$$H_{tot} = K_i + K_e + U_i + U_e + V_{ei}, \qquad (4)$$

with $K_{i,e}$ being the kinetic energy of ions or electrons indicated by the index; $U_{i,e}$ is the potential energy of interaction within the ionic or the electronic subsystem; $V_{ei}$ is the potential energy of interaction between two subsystems – the ionic and electronic ensembles – correspondingly split in half between them (as suggested in Ref. [27] and discussed below); $T_{i,e}$ are the temperatures in the ionic and electronic ensembles, and $f_{e,i}^{eq}$ are their partial equilibrium distribution functions (with $A$ being the normalization constant).

Since we are seeking to define the atomic temperatures, we consider an arbitrary vector field $\boldsymbol{B}(\vec{R},\vec{P})$ dependent on the ionic coordinates and momenta only, not the electronic system parameters. Applying to it the nabla operator with the distribution function from (3), analogously to Eqs.(1,2), we obtain the following equality linking the generalized electronic and ionic temperatures for arbitrary distribution functions of the ionic and electronic ensembles $f_i$ and $f_e$:

$$\int_\Omega \boldsymbol{B}(\vec{R}, \vec{P}) \cdot \left[ \frac{\boldsymbol{\nabla}\left[K_i + U_i + \frac{1}{2}V_{ei}\right]}{T_i} + \frac{\boldsymbol{\nabla}\left[K_e + U_e + \frac{1}{2}V_{ei}\right]}{T_e} \right] f_i f_e d\vec{R} d\vec{P} d\vec{r} d\vec{p} \qquad (5)$$
$$= \langle \boldsymbol{\nabla} \cdot \boldsymbol{B}(\vec{R}, \vec{P}) \rangle.$$

Where the capital letters $\vec{R}$ and $\vec{P}$ are used to denote the ionic coordinates and momenta, and small $\vec{r}$ and $\vec{p}$ are for the electronic ones.

Now, we may choose to use the nabla operator in the momentum or the configurational space of the atomic ensemble. For the atomic momentum space, $\boldsymbol{\nabla}_{\vec{P}}$, expression (5) reduces to

$$\frac{\langle \boldsymbol{B}(\vec{R}, \vec{P}) \cdot \boldsymbol{\nabla}_{\vec{P}} K_i \rangle}{T_i} = \langle \boldsymbol{\nabla}_{\vec{P}} \cdot \boldsymbol{B}(\vec{R}, \vec{P}) \rangle, \qquad (6)$$

assuming that all the interaction potentials are independent of the momenta (e.g., no magnetic force considered). Thus, the generalized atomic temperatures in the momentum space may be defined in the two-temperature state *via* the standard expression:

$$T_i = \frac{\langle \boldsymbol{B}(\vec{R}, \vec{P}) \cdot \boldsymbol{\nabla}_{\vec{P}} H_{tot} \rangle}{\langle \boldsymbol{\nabla}_{\vec{P}} \cdot \boldsymbol{B}(\vec{R}, \vec{P}) \rangle}. \qquad (7)$$

For the atomic temperatures in the configurational space, the situation is different because the electron-ion interaction depends on both, the atomic (ionic) and electronic parameters:

$$\frac{\langle \boldsymbol{B}(\vec{R}, \vec{P}) \cdot \boldsymbol{\nabla}_{\vec{R}}\left[U_i + \frac{1}{2}V_{ei}\right] \rangle}{T_i} + \frac{\langle \boldsymbol{B}(\vec{R}, \vec{P}) \cdot \boldsymbol{\nabla}_{\vec{R}}\left[\frac{1}{2}V_{ei}\right] \rangle}{T_e} = \langle \boldsymbol{\nabla}_{\vec{R}} \cdot \boldsymbol{B}(\vec{R}, \vec{P}) \rangle. \qquad (8)$$

As is typical for *ab-initio* MD simulations, or electronic-temperature dependent interatomic potentials, the electronic temperature is an external quantity for the atomic potential, and may be considered as a known parameter [28]. It, thus, follows from Eq.(8) that the configurational-space atomic temperatures in a two-temperature system may be expressed as follows:

$$T_i = \frac{\langle \boldsymbol{B}(\vec{R}, \vec{P}) \cdot \boldsymbol{\nabla}_{\vec{R}}\left[U_i + \frac{1}{2}V_{ei}\right] \rangle}{\langle \boldsymbol{\nabla}_{\vec{R}} \cdot \boldsymbol{B}(\vec{R}, \vec{P}) \rangle - \frac{1}{2T_e}\langle \boldsymbol{B}(\vec{R}, \vec{P}) \cdot \boldsymbol{\nabla}_{\vec{R}} V_{ei} \rangle}. \qquad (9)$$

This expression can be used in two-temperature simulations to evaluate atomic configurational-space temperatures.

As a consistency check, we may set $T_e \equiv T_i$ which then reduces Eq.(8) to

$$\frac{\langle \boldsymbol{B}(\vec{R}, \vec{P}) \cdot \boldsymbol{\nabla}_{\vec{R}}\left[U_i + \frac{1}{2}V_{ei} + \frac{1}{2}V_{ei}\right] \rangle}{T_i} = \langle \boldsymbol{\nabla}_{\vec{R}} \cdot \boldsymbol{B}(\vec{R}, \vec{P}) \rangle, \qquad (10)$$

and the ionic temperature restores the one-component limit:

$$T_i^{(1)} = \frac{\langle \boldsymbol{B}(\vec{R},\vec{P}) \cdot \boldsymbol{\nabla}_{\vec{R}} H_{tot} \rangle}{\langle \boldsymbol{\nabla}_{\vec{R}} \cdot \boldsymbol{B}(\vec{R},\vec{P}) \rangle}. \tag{11}$$

Let us emphasize that, strictly speaking, the condition $T_e \equiv T_i$ is not equivalent to the electron-ion equilibrium in a realistic simulation, because even if average values of the temperatures are equal, they still may fluctuate in different ways. Thus, the definition (9) must be used even in equilibrium two-temperature simulations instead of (11). Eq.(11) is only valid if the electronic temperature is *identically* equal to the ionic one at each simulation timestep.

We may also note that Eq.(9) has a divergency at $T_e \to 0$. In this case, one must consider in Eq.(8) that the electronic distribution function, entering the averaging brackets, is a Heaviside step function in the energy space at $T_e \to 0$. Thus, the spatial derivative can be expressed as

$$\boldsymbol{\nabla}_{\vec{R}} f_e = \delta(E_e - E_f) \frac{dE_e}{d\vec{R}} = \delta(E_e - E_f) \boldsymbol{\nabla}_{\vec{R}} V_{ei}(E_f)$$

where $E_f$ is the Fermi energy of the electrons and $E_e$ is single-electron energy. Albeit the quantity $\boldsymbol{\nabla}_{\vec{R}} V_{ei}(E_f)$ may be non-trivial to obtain, it allows us to formally get rid of the divergence in the definition of the atomic configurational temperature in the limit of cold electrons ($T_e \to 0$):

$$T_i = \frac{\langle \boldsymbol{B}(\vec{R},\vec{P}) \cdot \boldsymbol{\nabla}_{\vec{R}} \left[ U_i + \frac{1}{2} V_{ei} \right] \rangle}{\langle \boldsymbol{\nabla}_{\vec{R}} \cdot \boldsymbol{B}(\vec{R},\vec{P}) \rangle - \langle \boldsymbol{B}(\vec{R},\vec{P}) \cdot \boldsymbol{\nabla}_{\vec{R}} V_{ei}(E_f) \delta(E - E_f) \rangle}. \tag{12}$$

In this work, we are interested in an electronic system highly excited by irradiation and will not encounter situations in which this limiting case would be needed.

### 3. Atomic temperatures in the momentum space

Temperatures, defined for the vector fields in the momentum space only, are the most familiar ones. For example, setting $\boldsymbol{B}(\vec{P}) = \vec{P}$, the nabla operator in Eq.(7) is the one over the momentum space, and we obtain the definition of the temperature *via* the average kinetic energy of particles in the ensemble (assuming no center-of-mass motion) [26]:

$$T_{kin} = \frac{2}{3} \langle E_{kin} \rangle. \tag{13}$$

Such a definition of temperature is referred to as the *kinetic temperature* [18]. The same equation may also be used to define projections of the kinetic temperature on various axes, e.g. cartesian coordinates: $T_{kin,\beta} = \langle m v_\beta^2 \rangle$, with $m$ being the mass of a particle, and $v_\beta$ are the projections of its velocity on $\beta$ = X, Y, or Z.

One may also use, e.g., the field $\boldsymbol{B}(\vec{P}) = P^2 \vec{P}$, in which case the definition (7) produces

$$T = \frac{2}{5} \frac{\langle E_{kin}^2 \rangle}{\langle E_{kin} \rangle}.$$

Using Eq.(13), it may conveniently be rewritten as follows:

$$T_{fluc} = \sqrt{2/3}\sqrt{\langle E_{kin}^2\rangle - \langle E_{kin}\rangle^2} = \sqrt{2/3}\,\delta E_{kin}. \tag{14}$$

The fluctuation in the kinetic energies is defined as $\delta E_{kin} = \sqrt{\langle E_{kin}^2\rangle - \langle E_{kin}\rangle^2}$. Thus, this definition (14) will be referred to as the *fluctuational temperature*.

Note that the same definitions may be obtained *via* the direct evaluation of the first and second moments of the kinetic energies with the Maxwellian distribution [29]. However, Eq.(7) allows one to use an arbitrary vector field, therefore being a more general and powerful method than the method of moments.

The kinetic and fluctuational temperatures are equal to the thermodynamic temperature if the distribution in the momentum space is equilibrium one, that is, for the Maxwell-Boltzmann distribution. Out of equilibrium, the two do not coincide and thus may be used as a criterion to assess thermalization in the momentum subspace in simulations that have access to kinetic energies of individual particles (such as molecular dynamics [9]) or the distribution in the momentum or kinetic-energy space (such as the Boltzmann equation [1,30]).

## 4. Atomic temperatures in configurational space

One may define various temperatures solely in the one-particle configurational space of the atomic ensemble. A standard vector field used for the definition of the temperature entirely in the configurational space is $\boldsymbol{B}(\vec{R}) = \boldsymbol{\nabla}_{\vec{R}} H_{tot}$ [18,24,26]. Eq.(9) produces the following atomic *configurational temperature* in the two-temperature state:

$$T_{config} = \frac{\langle \boldsymbol{\nabla}_{\vec{R}} H_{tot} \cdot \boldsymbol{\nabla}_{\vec{R}}\left[U_i + \frac{1}{2}V_{ei}\right]\rangle}{\langle \boldsymbol{\nabla}_{\vec{R}}^2 H_{tot}\rangle - \frac{1}{2T_e}\langle \boldsymbol{\nabla}_{\vec{R}} H_{tot} \cdot \boldsymbol{\nabla}_{\vec{R}} V_{ei}\rangle} = -\frac{\langle \vec{F}\cdot \left[\vec{F}_i + \frac{1}{2}\vec{F}_{ie}\right]\rangle}{\langle \boldsymbol{\nabla}_{\vec{R}}\cdot \vec{F}\rangle + \frac{1}{2T_e}\langle \vec{F}\cdot \vec{F}_{ie}\rangle}, \tag{15}$$

where $\vec{F}_i$ is the ion-ion contribution to the force acting on an ion/atom; $\vec{F}_{ie}$ is the electronic contribution to the force; and $\vec{F} = \vec{F}_i + \vec{F}_{ie}$ is the total force acting on the atom.

In the case of $T_e \equiv T_i$, Eq.(15) reduces to the standard definition of the configurational temperature of a one-component system [18,26]:

$$T_{config}^{(1)} = \frac{\langle (\boldsymbol{\nabla}_{\vec{R}} H)^2\rangle}{\langle \boldsymbol{\nabla}_{\vec{R}}^2 H\rangle}, \tag{16}$$

A second definition of the atomic configurational temperature is needed if we want to trace thermalization in the configurational subspace. Analogously to the fluctuational temperature in the momentum space, we may introduce the field $\boldsymbol{B}(\vec{R}) = F^2\vec{F}$. This field does not produce an expression connected to fluctuations, but nonetheless, Eq.(10) gives a different definition of the temperature in the atomic configurational space:

$$T_{hyperconf} = -\frac{\langle (F^2\vec{F}) \cdot [\vec{F}_i + \frac{1}{2}\vec{F}_{ie}] \rangle}{\langle \nabla_{\vec{R}} \cdot (F^2\vec{F}) \rangle + \frac{1}{2T_e}\langle (F^2\vec{F}) \cdot \vec{F}_{ie} \rangle}, \tag{17}$$

where

$$\langle \nabla_{\vec{R}} \cdot (F^2\vec{F}) \rangle = (3F_x^2 + F_z^2 + F_z^2)\frac{\partial F_x}{\partial x} + (F_x^2 + 3F_z^2 + F_z^2)\frac{\partial F_y}{\partial y} + (F_x^2 + F_z^2 + 3F_z^2)\frac{\partial F_z}{\partial z}.$$

A definition of the temperature based on the field $|\boldsymbol{B}(\vec{R})| \sim F^s$ (for an arbitrary degree $s$) was referred to as *hyperconfigurational temperature* in Refs. [18,31]. Hyperconfigurational temperature, Eq.(17), may be compared to the configurational temperature from Eq.(15) to assess thermalization in the configurational subspace.

Note that, despite a rather cumbersome expression, Eq.(17) contains the same components as Eq.(15) – once calculated in a simulation, the partial forces and the force derivatives may be reused to easily construct the hyperconfigurational temperature.

In the case of $T_e \equiv T_i$, Eq.(17) reduces to the hypervirial-like expression from [18]:

$$T_{hyperconf}^{(1)} = -\frac{\langle F^4 \rangle}{\langle \nabla_{\vec{R}} \cdot (F^2\vec{F}) \rangle}$$

As an aside, we may consider another vector field $\boldsymbol{B}(\vec{R}) = \vec{R}$ [26]. Then the temperature definition (9) produces the following expression for the two-component *virial temperature*:

$$T_{vir} = -\frac{\langle \vec{R} \cdot [\vec{F}_i + \frac{1}{2}\vec{F}_{ie}] \rangle}{3 + \frac{1}{2T_e}\langle \vec{R} \cdot \vec{F}_{ie} \rangle}. \tag{18}$$

Eq.(18) for $T_e \equiv T_i$ reduces to the standard one-component virial expression [26]:

$$T_{vir}^{(1)} = -\frac{1}{3}\langle \vec{R} \cdot \vec{F} \rangle, \tag{19}$$

As was pointed out in Ref. [26], such a definition (19) (as well as the two-temperature definition (18)) is not periodic, and cannot be used in practice in common molecular dynamics simulations of solids or liquids with periodic boundary conditions. However, in the case of $T_e \equiv T_i$, this issue can be circumvented noticing that Eq.(19) functionally coincides with the definition of configurational pressure [32,33]:

$$P_{conf} = \frac{1}{3V}\langle \vec{R} \cdot \vec{F} \rangle,$$

where $V$ is the volume of the simulation box.

There are well-known methods of evaluation of the configurational pressure in systems with periodic boundary conditions which are often readily available, e.g. [34,35]. Thus, the one-component virial temperature, recast in terms of the configurational pressure, may use the same methods of evaluation:

$$T_{vir}^{(1)} = -P_{conf}V. \tag{20}$$

A comparison between the momentum-space and the configurational-space atomic temperatures will enable us to evaluate complete thermalization in the system [18]. For example, note that the coincidence of the kinetic temperature (Eq.(13)) with the virial temperature (Eq.(19)) is the equipartition theorem valid only in thermal equilibrium [18,26].

### III. Simulations

To study various thermalization processes in the nonequilibrium atomic system initiated by the ultrafast excitation of the electronic system of a solid by laser irradiation, we employ the XTANT-3 simulation toolkit [36,37]. It is a hybrid model consisting of the combined transport Monte-Carlo (for fast electrons) and Boltzmann collision integrals (for slow electrons) to trace the nonequilibrium evolution of the electronic system and the tight-binding molecular dynamics to follow atomic trajectories on the changing potential energy surface. All the details of the simulation tool may be found in Ref. [37]; here, we will only briefly recall the main points of the hybrid approach relevant to the current study.

The irradiation with an ultrafast laser pulse first drives the electronic system out of equilibrium [1,38]. Electrons absorb photons from the laser pulse, promoting them to higher energy states of a solid. In XTANT-3, electrons with energies above a chosen cut-off are simulated with the transport Monte Carlo (MC) method, tracing secondary electron cascades with event-by-event simulation technique, together with decays of core-shell holes (if any are excited by the photons or electron impact ionization) [37,38]. The inelastic scattering of electrons is traced with the linear response theory, applying the Ritchie-Howie model for the scattering cross-section [39,40]. The elastic scattering on target atoms is modeled with the screened Rutherford scattering cross section with the modified Molier screening parameter [40].

The evolution of the distribution function ($f_e(\varepsilon_i, t)$) of the electrons populating the valence and the conduction band of the material below the MC cut-off is calculated with the Boltzmann collisions integrals (BCI), including the electron-ion energy exchange ($I_{e-i}$), and electron-electron scattering resulting in the local thermalization in the electronic system ($I_{e-e}$) [28,37]:

$$\frac{df_e(\varepsilon_i, t)}{dt} = I_{e-e} + I_{e-i} + I_{MC},$$

$$I_{e-e} = -\frac{f_e(\varepsilon_i, t) - f_{eq}(\varepsilon_i, \mu, T_e, t)}{\tau_{e-e}}, \tag{21}$$

here, the distribution function describes fractional electronic populations on the energy levels $\varepsilon_i = \langle i|H_{TB}|i\rangle$, which are the eigenfunctions of the Hamiltonian at the current MD timestep; $I_{MC}$ is the source term describing the change of the distribution function due to photoabsorption, Auger-decays involving valence/conduction band, high-energy electrons scattering and influx calculated with the MC module; $\tau_{e-e}$ is the characteristic electron-electron relaxation time defining the electron-electron thermalization; $f_{eq}(\varepsilon_i, \mu, T_e, t)$ is the equivalent equilibrium Fermi-Dirac distribution with the same total number of (low-energy) electrons and energy content as in the transient nonequilibrium distribution [28]. The electron-ion coupling matrix element entering $I_{e-i}$ is evaluated from the tight-binding Hamiltonian [41]. The effects of the *electronic* non-equilibrium were studied elsewhere [28]; here, the main interest of the work is in the equilibration in the *atomic* system. Thus, we assume an instantaneous thermalization ($\tau_{e-e} \to 0$) which results in electrons adhering to the Fermi-Dirac distribution at all times, but with a temperature different from the atomic ones, defined by the energy transiently stored in the electronic system.

The atomic system is followed with the help of the molecular dynamics (MD) simulation in a box with periodic boundary conditions. Martyna-Tuckerman's 4[th]-order algorithm is used to propagate atomic trajectories [42]. Forces acting on atoms are evaluated with the help of the transferrable tight binding (TB) method [43–45]. For modeling of carbon-based materials, the TB parameterization from Ref. [43] is used; for silicon, the one from Ref. [44]. We use the MD timestep of 0.1 fs.

The transferable tight binding formalism provides the core-core repulsive terms for the ionic contribution to the force, $\vec{F}_i = -\mathbf{\nabla}_{\vec{R}} E_{rep}$, and the band energy for the electron-ion contribution to the force $\vec{F}_{ie} = -\sum f_e \langle i | \mathbf{\nabla}_{\vec{R}} H_{TB} | i \rangle$ [46], to be applied in Eqs.(15) and (17).

The TBMD method involves the calculation of the transient electronic energy levels (band structure, eigenstates of the transient electronic Hamiltonian) evolving in time, dependent on the positions of all the atoms in the simulation box and the transient electronic distribution function [45]. Thus, exciting the electronic system affects the interatomic potential and may lead to phase transitions – even without significant atomic heating, merely due to changes in the atomic interaction. This effect is known as the nonthermal melting [47,48].

Energy, transferred to atoms at each time-step *via* electron scattering on atoms (both, elastic scattering in the MC fraction or electrons, and electron-ion coupling in the BCI fraction), is delivered to atoms using the velocity scaling method [41]. This effect is known as the non-adiabatic electron-ion coupling or electron-phonon coupling [2].

In contrast to the standard two-temperature molecular dynamics (see, e.g., [5,49]), XTANT-3 relies on the tight-binding formalism to calculate the collective atomic potential energy surface dependent on the transient state of the electronic ensemble. It includes the evolution of the electronic energy levels (density of states). The electron-ion (electron-phonon) coupling is calculated on-the-fly depending on the state of the electronic and atomic systems. The formalism is capable of describing electronic non-equilibrium if required in a simulation of irradiation. The combined approach developed in XTANT-3 allows us to trace all the essential effects of irradiation on the matter, in a reasonable agreement with experiments (see e.g. [38,50–52]). Most importantly for this work, it enables tracing nonequilibrium processes in the system. Various definitions of temperatures (listed in Section II) were implemented in XTANT-3 to trace the behavior of the nonequilibrium temperatures induced by the irradiation, including thermal and nonthermal effects under various conditions.

## 1. Thermalization

We start with the test case of assessing the thermalization of the atomic system in a simulation box without additional external perturbances. For this illustration, we use diamond with 216 atoms in the supercell. The simulation box is set in the minimum of the potential energy *via* the steepest descent method [9,37]. Then, all the atoms are given equal velocities $v_0 = \sqrt{2T/M}$, corresponding to the value of room temperature (*T*=300 K), with directions randomly uniformly distributed in the solid angle.

Obviously, such a distribution of atomic velocities does not correspond to the Maxwellian distribution in the momentum space, and thus the initial nonequilibrium may be expected to be observable in the simulation. Indeed, Figure 1 shows that the fluctuational temperature at the beginning of the simulation is near zero, whereas the kinetic temperature is defined by $v_0$ (double the room temperature). However, very quickly the kinetic and fluctuational temperature merge by the time of ~30 fs, see inset in Figure 1. It demonstrates that the equilibration in the momentum space occurs after just a few atomic oscillations, within the characteristic time of optical phonon vibrations.

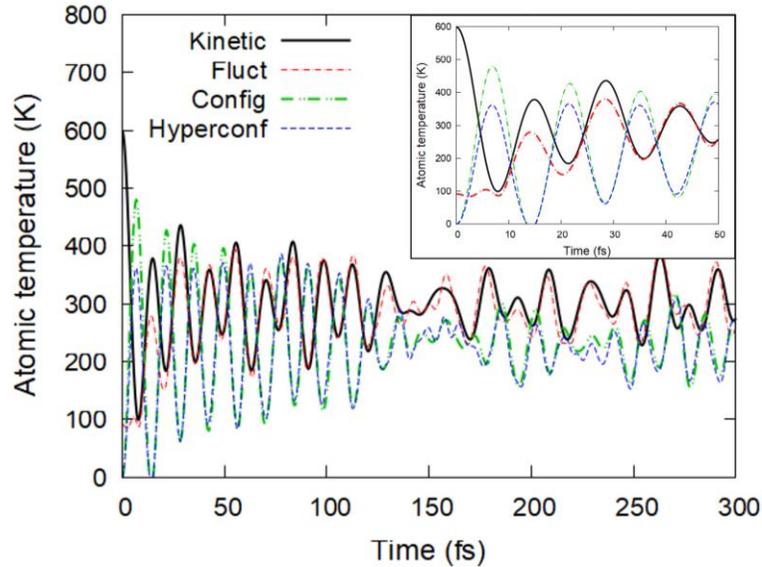

*Figure 1. Kinetic (solid line), fluctuational (dash-dotted), configurational (dash-dot-dotted) and hyperconfigurational (dashed) atomic temperatures in diamond, 216 atoms in the simulation box with periodic boundary conditions, simulated with XTANT-3 (inset zooms on the first 50 fs, emphasizing equilibration of the kinetic and fluctuational temperatures).*

In the configurational space, the configurational and hyperconfigurational atomic temperatures are very close from the very beginning of the simulation, apart from different heights of the peaks in $T_{config}$. By the time of ~40-50 fs, the two temperatures in the configurational space are nearly indistinguishable. It indicates that, at least in the relatively small simulation box considered here, the system is equilibrated in the real space.

The momentum-space (both, $T_{kin}$ and $T_{fluc}$) and the configurational-space ($T_{config}$ and $T_{hyperconf}$) temperatures oscillate around the room temperature after ~150 fs, indicating the realization of the equipartition of energy, and, thus, complete thermalization in the system. By that time, the oscillations in all temperatures become less regular, demonstrating the loss of artificial coherence introduced by the initial conditions. We emphasize that the equilibration in each subspace of the phase space – the momentum and the configurational – takes shorter times than the equilibration between the two.

The results validate the derivation of various atomic temperatures in the two-temperature state based on the factorized Gibbs distribution, Eq. (3). Such definitions are consistent, and lead to complete thermalization in the atomic ensemble.

## 2. Laser irradiation

Let us now proceed to the simulation of the response of diamond and silicon to laser irradiation. We start with the simulation of nonthermal graphitization of diamond irradiated with a 10 fs (full width at half maximum, FWHM) laser pulse with the deposited dose of 1.5 eV/atom. At such conditions, diamond undergoes a phase transition to an overdense graphite-like state (a mix of sp2 and sp3 carbon) within ~100-200 fs, see the atomic snapshots in Appendix and details in Refs. [53–55]. Let us zoom at a small section of diamond, consisting of 64 atoms in a simulation box, which undergoes a complete transition to graphite (in a larger simulation box, a few nucleation centers usually form, each of them creating its own graphite-like structure [50,56], which will only obscure the presentation of the results here). In the transition from the cubic diamond structure to the highly anisotropic structure of the parallel graphitic planes, the temperatures behave as follows, see Figure 2.

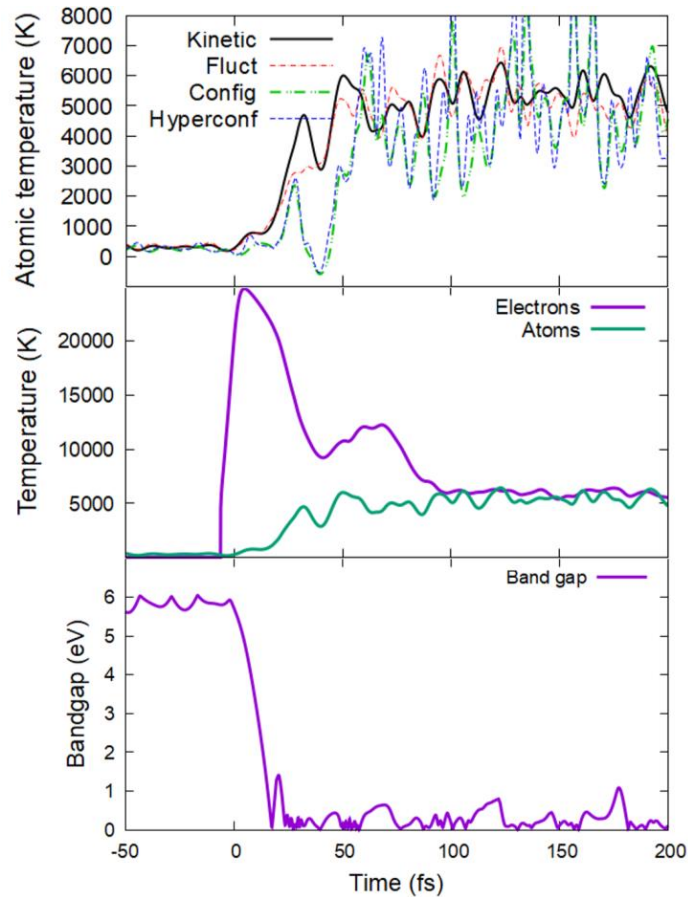

*Figure 2. Kinetic (solid line), fluctuational (thin dash-dotted), configurational (dash-dot-dotted) and hyperconfigurational (thin dashed) atomic temperatures in diamond irradiated with the 10 fs FWHM laser pulse centered around 0 fs, absorbed dose of 1.5 eV/atom, simulated with XTANT-3. (Middle panel) Electronic and kinetic atomic temperatures. (Bottom panel) Band gap in the same simulation.*

During the ultrafast phase transition, both of the momentum-space temperatures – kinetic and fluctuational – stay close, except for the short period between 20 fs and 50 fs, when the phase transition occurs. As was studied in detail before [53–55], the nonthermal graphitization of diamond is triggered by the bandgap collapse, which can be seen in Figure 2 (bottom panel). When the band gap collapses, indicating a coherent transition to a semimetallic graphite-like state (see Appendix), the atomic system disequilibrates transiently. Then, the equilibrium *in the momentum subspace* takes place by the time of ~50 fs. After that, the two momentum-space temperatures stay close, oscillating around the same value.

In contrast, configurational and hyperconfigurational temperatures depart from the momentum-space ones. First, both configurational-space temperatures drop down slightly during the laser pulse (around 0 fs), while the momentum-space temperatures rise. This drop takes place due to modification of the interatomic potential induced by the excitation of electrons. The electrons define the interatomic potential, thus, changes in the electronic distribution function (excitation by the laser pulse) directly affect the potential energy and interatomic forces [38,53,57]. At the same time, there is a slight divergence between the configurational and hyperconfigurational temperatures, which quickly equilibrate again.

Except for this short period around ~0 fs (during the laser pulse excitation of the electronic system), the atomic system is in equilibrium *in the configurational subspace* during the entire simulation. However, the difference between the momentum-space and configurational-space temperatures persists from the laser pulse arrival up to the completion of the nonthermal phase transition (by the time of ~100-150 fs). The equilibration between the electronic temperature and the atomic (kinetic) one also takes place within the same time window (middle panel in Figure 2).

Note that the second drop in the configurational-space temperatures takes place around the time of the onset of the phase transition, ~20 fs. Interestingly, the configurational-space temperatures during and after the electronic excitation with the laser pulse transiently turn negative (the minimum around ~40 fs). Despite the ongoing debate on the possibility of negative absolute temperatures [58–60], we emphasize that our system is out-of-equilibrium: atomic momentum-space temperatures are positive, and only the configurational-space temperatures are negative during the ongoing phase transition. As could be seen in Section II, negative configurational-space temperatures are allowed by definition, as they are connected with the forces (or microscopic stresses) in the system.

The negative configuration-space temperatures appear due to modifications of the interatomic potential because of the excitation of the electronic system (note the extremely high electronic temperatures reached during this time, Figure 2). Excited electronic higher-energy states are typically anti-bonding [61]. Thus, the character of the atomic potential transiently changes upon ultrafast electronic excitation: the potential-energy minimum shifts to larger values, forming positive microscopic pressures attempting to expand the material. Changes in the forces sign, from attractive to repulsive, produce negative configurational temperatures lasting only until the atoms move to their new positions corresponding to the new potential minimum.

Next, we simulate silicon irradiated with a 10 fs (FWHM) laser pulse and the deposited dose of 1.2 eV/atom. This dose is above the nonthermal melting leading to the high-density liquid silicon (the threshold is ~0.9 eV/atom [62]). Various temperatures in silicon in this simulation are shown in Figure 3.

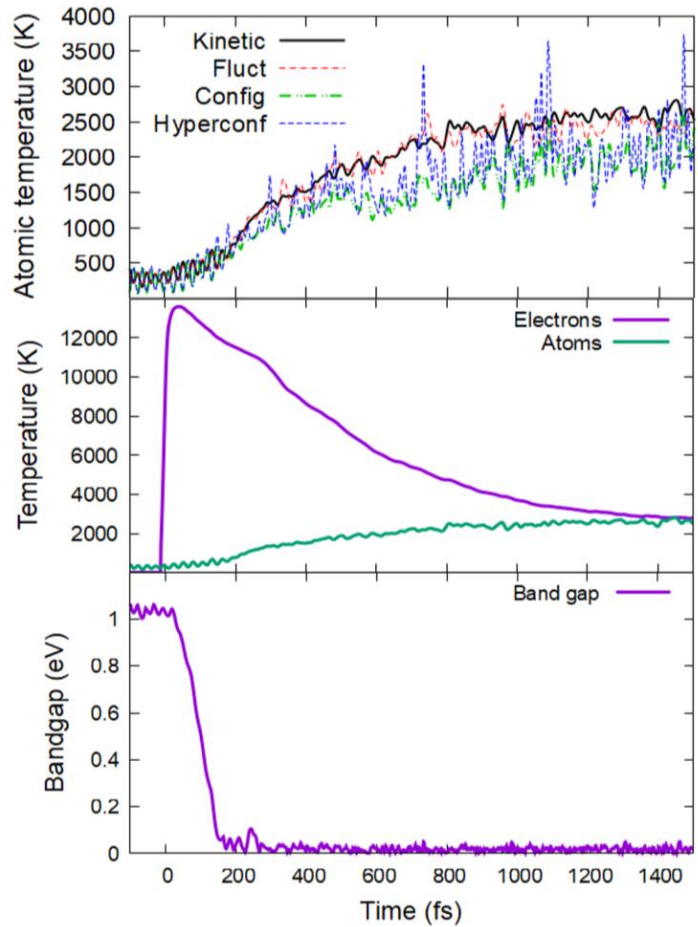

*Figure 3. (Top panel) Kinetic (solid line), fluctuational (thin dash-dotted), configurational (dash-dot-dotted) and hyperconfigurational (thin dashed) atomic temperatures in silicon irradiated with the 10 fs FWHM laser pulse centered around 0 fs, absorbed dose of 1.2 eV/atom, simulated with XTANT-3. (Middle panel) Electronic and kinetic atomic temperatures. (Bottom panel) Band gap in the same simulation.*

In silicon, the nonthermal phase transition is not as fast as in diamond, onsetting at the time of ~150-200 fs (see Appendix), and taking place up to some 500-700 fs [38,62]. During the entire simulation time, the momentum-space temperatures stay close to each other. The two configurational-space temperatures are also close to each other, but different from the momentum-ones.

Similarly to the diamond case above, the configurational-space temperatures in silicon start to deviate from the momentum-space ones during the laser pulse excitation of the electrons (see the rise of the electronic temperature, middle panel in Figure 3), but the deviation is barely noticeable until the onset of the phase transition. The drop of the configurational-space temperatures related to changes in the interatomic potential is much smaller, which is consistent with the slower phase transition in silicon than in diamond [38].

The onset of the atomic disorder can be seen by the collapse of the bandgap (bottom panel in Figure 3, showing a transition into a liquid metallic state, see [38,62] for details). After that time of ~200 fs, the configurational-space temperatures remain below the momentum-space ones until the end of the simulation. It indicates that the phase transition in Si is a nonequilibrium process, indicated by non-matching various atomic temperatures.

We note again that during the entire simulation, including the laser-irradiation stage, electron-ion nonequilibrium, and the phase transition itself, the kinetic and the fluctuational atomic temperatures stay close, with only minor deviations, showing that the system remains equilibrated in the momentum subspace. The configurational-space temperatures are also close to one another but are different from the momentum-space ones. It shows that the complete thermal equilibrium is not achieved until much later times after the phase transition, related to mechanical relaxation (relaxation of pressure/stresses). That means that the transient bi-temperature state forms in the atomic system of irradiated material after ultrafast material excitation: partial equilibrium in the momentum subspace, and partial equilibrium in the configurational subspace with its own temperature (apart from the third, different electronic temperature).

## IV. Discussion

Our results suggest that in many cases (apart from a short-lived initial state), the nonequilibrium atomic system of a target under external perturbation, such as laser or swift heavy ion irradiation, may be described with only two parameters: for example, the kinetic temperature and the configurational temperature, describing the separate Gibbs distributions in the atomic momentum and configurational subspaces, respectively. Such a multi-temperature state is much simpler than fully nonequilibrium distribution, and further development of theory and simulations may use this fact to their advantage.

As an example, let us consider a multi-temperature state in which each term in the Hamiltonian (4) corresponds to its own temperature. In this rather general case, the factorized Gibbs distribution function can be written as:

$$f(r,p) = A\, e^{-K_e/T_e^K} e^{-U_e/T_e^U} e^{-K_i/T_i^K} e^{-U_i/T_i^U} e^{-V_{ei}/T_{ei}}, \qquad (22)$$

where the upper indices mark the subspace the temperature belongs to: $K$ is for the momentum space (according to the kinetic energy), and $U$ is for the configurational space (related to the potential energy); the lower indices mark the ensemble: electrons (*e*) or ions (*i*); and a separate electron-ion interaction term (*ei*) related to their interaction $V_{ei}$ and the corresponding temperature $T_{ei}$

For simplicity, we are referring here to all the momentum-space temperatures as the kinetic temperature; and all the configurational-space temperatures are referred to as the configurational temperature. The momentum-space partial distributions are the Maxwell-Boltzmann distribution functions but with their own temperatures, different from the configurational-space partial distributions for each ensemble, and possibly from the interaction temperature.

Using the standard technique of deriving the transport equations for an arbitrary time-independent quantity, $\varphi$ (see, e.g., Refs. [33,63]), we multiply the Liouville equation with each energy term separately, and integrate it over the local phase space (allowing for global spatial gradients to trace global energy flows):

$$\int_\Omega \frac{df}{dt}\varphi d\Omega + \int_\Omega \sum_{\gamma=1}^{\nu}\sum_{k=1}^{N_\gamma}\left\{\vec{v}_{\gamma k}\frac{\partial f}{\partial \vec{r}_{\gamma k}}\varphi - \frac{\partial H_{tot}}{\partial \vec{r}_{\gamma k}}\frac{\partial f}{\partial \vec{p}_{\gamma k}}\varphi\right\}d\Omega = 0, \quad (23)$$

where the index $k$ indicates summation over various sorts of particles $\gamma$ in the ensemble of $N_\gamma$ particles, $\nu$ stands for the number of different sorts, integrated over the phase space $\Omega$. Here, we assume no external forces, and no hydrodynamical motion (no center of mass motion in each local volume).

In our case, we consider two different sorts of particles – electrons and ions – resulting in the general equation:

$$\int_\Omega \frac{df}{dt}\varphi d\Omega + \int_\Omega \left\{\left[\vec{v}_e\frac{\partial f}{\partial \vec{r}_e}\varphi - \frac{\partial H_{tot}}{\partial \vec{r}_e}\frac{\partial f}{\partial \vec{p}_e}\varphi\right] + \left[\vec{v}_i\frac{\partial f}{\partial \vec{r}_i}\varphi - \frac{\partial H_{tot}}{\partial \vec{r}_i}\frac{\partial f}{\partial \vec{p}_i}\varphi\right]\right\}d\Omega = 0. \quad (24)$$

Now, we use Eq.(24) to derive the equations of transport of the five energies (temperatures) from Eq.(21).

For example, using the kinetic energy of electrons, $\varphi = K_e$, the time derivative produces:

$$\begin{aligned}\int_\Omega \frac{df}{dt}K_e d\Omega &= \frac{\partial}{\partial t}\langle K_e\rangle - \langle\frac{\partial K_e}{\partial t}\rangle \\ &= \frac{\partial\langle K_e\rangle}{\partial T_e^K}\frac{\partial T_e^K}{\partial t} + \frac{\partial\langle K_e\rangle}{\partial T_i^K}\frac{\partial T_i^K}{\partial t} + \frac{\partial\langle K_e\rangle}{\partial T_e^U}\frac{\partial T_e^U}{\partial t} + \frac{\partial\langle K_e\rangle}{\partial T_i^U}\frac{\partial T_i^U}{\partial t} + \frac{\partial\langle K_e\rangle}{\partial T_{ei}}\frac{\partial T_{ei}}{\partial t} \\ &= \frac{\partial\langle K_e\rangle}{\partial T_e^K}\frac{\partial T_e^K}{\partial t},\end{aligned} \quad (25)$$

where only one term is left, while all the other terms vanish because the electronic kinetic energy does not depend on other temperatures, as per definition (22).

The spatial derivative term, integrated by parts and applying Leibniz integral rule, is:

$$\begin{aligned}\int_\Omega \left\{\vec{v}_e\frac{\partial f}{\partial \vec{r}_e}K_e + \vec{v}_i\frac{\partial f}{\partial \vec{r}_i}K_e\right\}d\Omega &= \frac{\partial}{\partial \vec{r}_e}\langle\vec{v}_e K_e\rangle - \langle\frac{\partial}{\partial \vec{r}_e}(\vec{v}_e K_e)\rangle + \frac{\partial}{\partial \vec{r}_i}\langle\vec{v}_i K_e\rangle - \langle\frac{\partial}{\partial \vec{r}_i}(\vec{v}_i K_e)\rangle \\ &= \frac{\partial}{\partial \vec{r}_e}\vec{q}_e^K,\end{aligned} \quad (26)$$

where we defined the electron kinetic energy current as $\vec{q}_e^K = \langle\vec{v}_e K_e\rangle$, and took into account that the electronic kinetic energy is independent of the coordinates, the electronic energies and ionic velocities are uncorrelated $\langle\vec{v}_i K_e\rangle = 0$, and the system has no center-of-mass flow in our consideration here.

The momentum derivative term is:

$$\int_\Omega \left\{ \frac{\partial H_{tot}}{\partial \vec{r}_e} \frac{\partial f}{\partial \vec{p}_e} K_e + \frac{\partial H_{tot}}{\partial \vec{r}_i} \frac{\partial f}{\partial \vec{p}_i} K_e \right\} d\Omega = -\left\langle \frac{\partial H_{tot}}{\partial \vec{r}_e} \frac{\partial K_e}{\partial \vec{p}_e} \right\rangle - \left\langle \frac{\partial H_{tot}}{\partial \vec{r}_i} \frac{\partial K_e}{\partial \vec{p}_i} \right\rangle \quad (27)$$
$$= \langle \vec{F}_{ee} \vec{v}_e \rangle + \langle \vec{F}_{ei} \vec{v}_e \rangle,$$

because the electronic kinetic energy is independent of the ionic momenta $\frac{\partial K_e}{\partial \vec{p}_i} = 0$, global momentum divergence vanishes [64], and the ion-ion interaction is independent of the electronic coordinate.

Analogously, we can evaluate the corresponding terms for the other four energies from Eq.(22). Combining them together, we obtain the system of equations for the five temperatures considered:

$$\begin{cases} \frac{\partial \langle K_e \rangle}{\partial T_e^K} \frac{\partial T_e^K}{\partial t} = -\frac{\partial}{\partial \vec{r}_e} \vec{q}_e^K + \langle \vec{F}_{ee} \vec{v}_e \rangle + \langle \vec{F}_{ei} \vec{v}_e \rangle \\ \frac{\partial \langle U_e \rangle}{\partial T_e^U} \frac{\partial T_e^U}{\partial t} = -\frac{\partial}{\partial \vec{r}_e} \vec{q}_e^U - \langle \vec{F}_{ee} \vec{v}_e \rangle \\ \frac{\partial \langle K_i \rangle}{\partial T_i^K} \frac{\partial T_i^K}{\partial t} = -\frac{\partial}{\partial \vec{r}_i} \vec{q}_i^K + \langle \vec{F}_{ii} \vec{v}_i \rangle - \langle \vec{F}_{ei} \vec{v}_i \rangle, \\ \frac{\partial \langle U_i \rangle}{\partial T_i^U} \frac{\partial T_i^U}{\partial t} = -\frac{\partial}{\partial r_i} \vec{q}_i^U - \langle \vec{F}_{ii} \vec{v}_i \rangle \\ \frac{\partial \langle V_{ei} \rangle}{\partial T_{ei}} \frac{\partial T_{ei}}{\partial t} = -\frac{\partial}{\partial \vec{r}} \vec{q}_{ei}^U - \langle \vec{F}_{ei} \vec{v}_e \rangle + \langle \vec{F}_{ei} \vec{v}_i \rangle \end{cases} \quad (28)$$

where the energy currents are defined for all terms: for the kinetic energy of ions $\vec{q}_e^K = \langle \vec{v}_e K_e \rangle$; for the potential energy of electrons $\vec{q}_e^U = \langle \vec{v}_e U_e \rangle$; the potential energy of ions $\vec{q}_i^U = \langle \vec{v}_i U_i \rangle$; and the potential energy of electron-ion interaction with the total spatial derivative used: $\frac{\partial}{\partial \vec{r}} q_{ei}^U = \frac{\partial}{\partial \vec{r}_e} \langle \vec{v}_e V_{ei} \rangle + \frac{\partial}{\partial \vec{r}_i} \langle \vec{v}_i V_{ei} \rangle$. The force-velocity correlators define the couplings between the various temperatures: $\langle \vec{F}_{ee} \vec{v}_e \rangle$ is responsible for equilibration between the kinetic and configuration temperatures in the electronic system; $\langle \vec{F}_{ii} \vec{v}_i \rangle$ equilibrates the kinetic and configuration temperatures in the ionic system; $\langle \vec{F}_{ei} \vec{v}_e \rangle$ is responsible for energy exchange between the electrons and the interaction term; and $\langle \vec{F}_{ei} \vec{v}_i \rangle$ is the energy exchange between the ions and the interaction term (taking into account that $\vec{F}_{ei} = -\vec{F}_{ie}$).

Now, let us consider how this system (28) simplifies in the case of four temperatures: when the interaction is not considered as a separate subsystem but only coupled electrons and ions are included. In this case, the electron-ion interaction term depends on the configurational electronic and ionic temperatures:

$$\frac{\partial \langle V_{ei} \rangle}{\partial T_{ei}} \frac{\partial T_{ei}}{\partial t} = \frac{\partial \langle V_{ei} \rangle}{\partial T_e^U} \frac{\partial T_e^U}{\partial t} + \frac{\partial \langle V_{ei} \rangle}{\partial T_i^U} \frac{\partial T_i^U}{\partial t} = -\frac{\partial}{\partial \vec{r}} \vec{q}_{ei}^U - \langle \vec{F}_{ei} \vec{v}_e \rangle + \langle \vec{F}_{ei} \vec{v}_i \rangle \quad (29)$$

The last equation in the system (28) is thus excluded and instead, Eq.(29) multiplied with the factor of 1/2 is added to the equations for the electronic and ionic configurations temperatures, resulting in:

$$\begin{cases} \dfrac{\partial \langle K_e \rangle}{\partial T_e^K} \dfrac{\partial T_e^K}{\partial t} = -\nabla \vec{q}_e^K + \langle \vec{F}_{ee} \vec{v}_e \rangle + \langle \vec{F}_{ei} \vec{v}_e \rangle \\ \dfrac{\partial \langle U_e + V_{ei}/2 \rangle}{\partial T_e^U} \dfrac{\partial T_e^U}{\partial t} + \dfrac{\partial \langle U_e + V_{ei}/2 \rangle}{\partial T_i^U} \dfrac{\partial T_i^U}{\partial t} = -\nabla \left( \vec{q}_e^U + \dfrac{1}{2} \vec{q}_{ei}^U \right) - \langle \vec{F}_{ee} \vec{v}_e \rangle - \dfrac{1}{2} \langle \vec{F}_{ei} \vec{v}_e \rangle + \dfrac{1}{2} \langle \vec{F}_{ei} \vec{v}_i \rangle \\ \dfrac{\partial \langle K_i \rangle}{\partial T_i^K} \dfrac{\partial T_i^K}{\partial t} = -\nabla \vec{q}_i^K + \langle \vec{F}_{ii} \vec{v}_i \rangle - \langle \vec{F}_{ei} \vec{v}_i \rangle \\ \dfrac{\partial \langle U_i + V_{ei}/2 \rangle}{\partial T_e^U} \dfrac{\partial T_e^U}{\partial t} + \dfrac{\partial \langle U_i + V_{ei}/2 \rangle}{\partial T_i^U} \dfrac{\partial T_i^U}{\partial t} = -\nabla \left( \vec{q}_i^U + \dfrac{1}{2} \vec{q}_{ei}^U \right) - \langle \vec{F}_{ii} \vec{v}_i \rangle - \dfrac{1}{2} \langle \vec{F}_{ei} \vec{v}_e \rangle + \dfrac{1}{2} \langle \vec{F}_{ei} \vec{v}_i \rangle \end{cases} \quad (30)$$

In this four-temperature model, the coupling is directly between the electrons and ions, not *via* a separate equation for the interaction terms.

As a consistency check, we now consider the two-temperature case, which assumes that each ensemble is locally thermalized: $T_e^K = T_e^U = T_e$ and $T_i^K = T_i^U = T_i$, each with its own thermodynamic temperature. In this case, the system (30) summed pairwise for electrons and ions reduces to the following:

$$\begin{cases} \dfrac{\partial \langle E_e \rangle}{\partial T_e} \dfrac{\partial T_e}{\partial t} + \dfrac{\partial \langle E_e \rangle}{\partial T_i} \dfrac{\partial T_i}{\partial t} = -\nabla \vec{q}_e + \langle \vec{F}_{ei} \dfrac{\vec{v}_e + \vec{v}_i}{2} \rangle \\ \dfrac{\partial \langle E_i \rangle}{\partial T_e} \dfrac{\partial T_e}{\partial t} + \dfrac{\partial \langle E_i \rangle}{\partial T_i} \dfrac{\partial T_i}{\partial t} = -\nabla \vec{q}_i - \langle \vec{F}_{ei} \dfrac{\vec{v}_e + \vec{v}_i}{2} \rangle \end{cases}, \quad (31)$$

with the total energies in each system marked as $E_{e,i} = K_{e,i} + U_{e,i} + V_{ei}/2$, and the total heat flows $\vec{q}_{e,i} = \vec{q}_{e,i}^K + \vec{q}_{e,i}^U + 1/2\vec{q}_{ei}^U$. Note that the coupling term, defined *via* equal splitting of the electron-ion interaction potential between the subsystems (cf. Eq.(3)), ensures that the energy loss by electrons is equal to the energy gain by ions (and vise versa).

Eq.(31) is analogous to the two-temperature model for the case of strongly coupled electrons and ions derived in Ref. [27], with the difference that here we assumed classical particles for both systems and thus arrived at the classical coupling term expressed as a velocity-force correlator $Z_{ei} = \langle \vec{F}_{ei} (\vec{v}_e + \vec{v}_i)/2 \rangle$, whereas Ref. [27] considered a quantum case.

This system of equations reduces to the well-known weakly-coupled limit of the two-temperature model (see e.g. Ref. [1,65,66]) if $\dfrac{\partial \langle E_e \rangle}{\partial T_i} = \dfrac{\partial \langle E_i \rangle}{\partial T_e} = 0$:

$$\begin{cases} C_e \dfrac{\partial T_e}{\partial t} = -\nabla \vec{q}_e + Z_{ei} \\ C_i \dfrac{\partial T_i}{\partial t} = -\nabla \vec{q}_i - Z_{ei} \end{cases}, \quad (32)$$

where the electronic and ionic heat capacities are $C_e = \dfrac{\partial \langle E_e \rangle}{\partial T_e}$ and $C_i = \dfrac{\partial \langle E_i \rangle}{\partial T_i}$.

Note that Eq.(31) accounts for the dependences of the ionic energy on the electronic temperature and vice versa, and thus, in principle, is capable of modeling of nonthermal effects associated with the changes of interatomic potential due to electronic excitation, while the weakly-coupled limit, Eq.(32), is not.

One may analogously derive all the multi-temperature thermodynamic equations including pressures, external forces, and hydrodynamic motion (following the methodology from Refs. [33,63] and using the five or four-temperature distribution). It should also be possible to derive the quantum version of the equations following the recipe from Ref. [27] (which should replace the classical correlators in Eq.(31) with appropriate quantum terms). Both tasks are, however, beyond the scope of the present work and are left for future dedicated research.

Last but not least, the question arises of how various atomic temperatures may be observed experimentally. Each temperature is associated with certain phase-space variables. As we discussed, the kinetic temperature of atoms is associated with the atomic momenta or velocities and thus may be measured with the methods having access to atomic momenta, such as, e.g., velocity-resolved pump-probe spectroscopy [67]. The configurational temperature, on the other hand, is defined by the potential energy of the atoms, and thus, we may assume, it can be accessed *via* methods sensitive to the interatomic potential and atomic distribution in real space, such as ultrafast X-ray or electron diffraction [68].

## V. Conclusions

Matter driven far out of equilibrium, for example *via* ultrafast laser irradiation, demonstrates unusual kinetics and properties. We studied here various atomic temperatures defined in the momentum and the configurational subspaces of the phase space in a "two-temperature state" – with an excited electronic system. Two definitions in each subspace were presented: kinetic and fluctuational temperatures in the momentum space of atoms, and configurational and hyperconfigurational temperatures in the real space. A coincidence within each pair of them indicates partial equilibration in the corresponding subspace; a coincidence between the different subspaces indicates complete thermalization of the atomic system.

We showed that the thermalization of the atomic system in the momentum space (establishment of the kinetic temperature) takes place extremely fast, within a few atomic oscillations (typically, a few tens of femtoseconds). Its local thermalization in the configurational space also occurs quickly but towards its own configurational temperature different from the kinetic one (both different from the third, electronic temperature in an irradiated system). The complete thermalization between the kinetic- and the configurational-space atomic temperatures requires longer times.

We discussed that this multi-temperature thermodynamic state of atoms allows us to write the factorized form of the atomic distribution function and a corresponding set of the energy balance equations, describing the atomic system not with a single parameter – the thermodynamic temperature – but with two: kinetic and configurational temperatures (while electrons are described with their own one or two temperatures). Allowing for the construction of a nonequilibrium thermodynamic description, this special case of nonequilibrium is significantly simpler than a completely nonequilibrium kinetic theory.

## VI. Data and code availability

The code XTANT-3 used to produce the data is available from [36].

## VII. Acknowledgments

NM thanks the financial support from the Czech Ministry of Education, Youth, and Sports (grant nr. LM2023068), and the computational resources provided by the e-INFRA CZ project (ID:90254), supported by the Ministry of Education, Youth and Sports of the Czech Republic.

## VIII. Appendix

Examples of the atomic snapshots in irradiated diamond and silicon, undergoing ultrafast nonthermal phase transitions are shown in Figure 4. Diamond forms a graphite-like (sp2-bonded) structure within the time window of ~150-200 fs, as was supported by experiments [50]. Silicon turns into high-density metallic liquid on the timescale of ~200-300 fs at the considered radiation dose [62]. Both phase transitions are triggered by changes in the interatomic potential, induced by the excitation of electrons which modifies interatomic forces, not due to atomic heating via electron-phonon coupling – nonthermal phase transitions (see details, e.g., in [38]). The nonthermal phase transitions are associated with the electron bandgap collapse, as discussed in the main text.

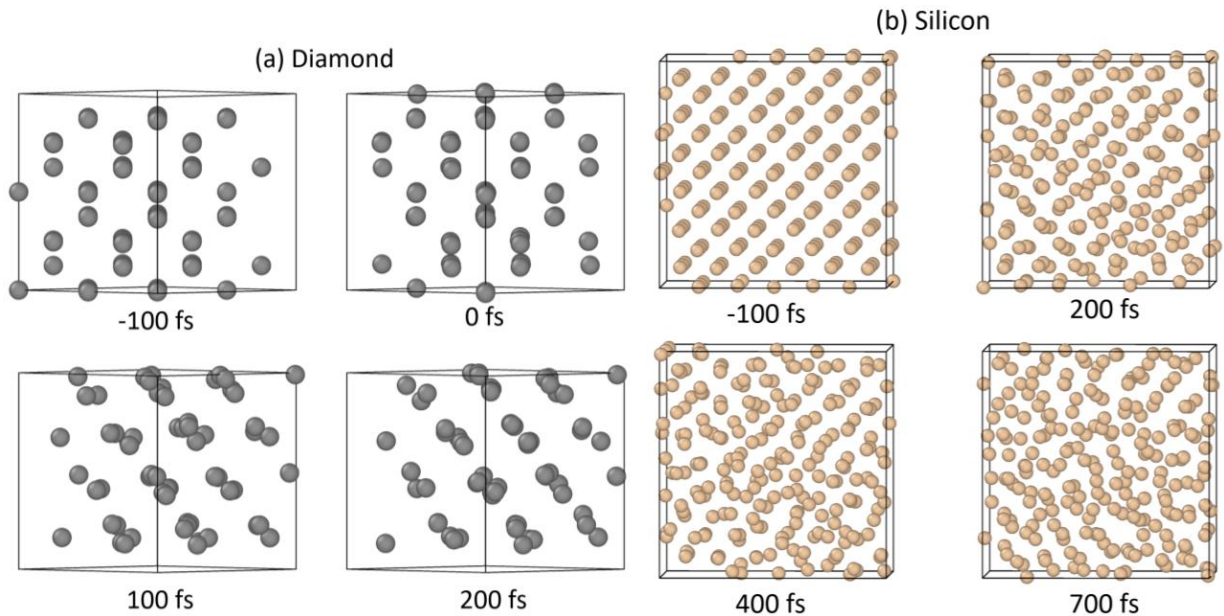

*Figure 4. (a) Diamond irradiated with 1 eV/atom. (b) Silicon irradiation with 1.2 eV/atoms.*